# Recent advances in spatial light modulator-based three-dimensional optical imaging (Invited)

Joseph Rosen

School of Electrical and Computer Engineering, Ben-Gurion University of the Negev, Beer-Sheva 8410501, Israel

rosenj@bgu.ac.il;

**ABSTRACT**

Phase-only spatial light modulators (SLMs) are used in optical systems for several purposes. In this article, the main landmarks of SLM-based imaging systems are surveyed. In addition to conventional two-dimensional imaging, these systems are useful for multidimensional imaging, axial sectioning, field-of-view expansion, improved image resolution, imaging through scatterers, and depth-of-field control. The SLMs in this review are positioned in the system aperture and modulate the input light in various ways to achieve different imaging goals. This review begins with the nearly 20-year-old Fresnel incoherent correlation holography system, continues with coded-aperture holography, and progresses to the most recent versions of interferenceless coded-aperture holography systems.

## 1. Introduction

Several review articles about advanced optical imaging methods have been published recently [1–8]. The present article reviews the field of optical imaging differently from the perspective of a component that has become essential in many imaging systems: the spatial light modulator (SLM) [9]. As shown below, SLMs have opened new possibilities and extended imaging capabilities, making it worthwhile to review the main landmarks in SLM-based imaging. Phase-only SLMs were introduced into optical systems in the late 1980s [10]. These devices have been applied primarily in optical correlators [11], free-space optical interconnections [12], and computer-generated holography [13]. In 2007, the phase-only SLM was first introduced into a holographic imaging system called Fresnel incoherent correlation holography (FINCH) [14]. FINCH is the result of approximately 60 years of extensive research in holography. Hence, a brief overview of holographic imaging and its main historical milestones can help readers understand the topics of this article.

The history and classification of holographic imaging are schematically illustrated in Fig. 1. The classification of holograms for imaging applications is split between spatially coherent and incoherent holograms, denoted by the two columns of the figure, where the state of coherence in this classification refers to the light emitted from the observed objects in the recording stage of the holograms. The other classification is between optical and digital holograms, as shown in the two rows of Fig. 1, which correspond to the means of image reconstruction from holograms. In the optical category, the developed hologram is usually illuminated by an optical source to display the reconstructed image directly to the observer's eyes, whereas in the digital case, the hologram is recorded by a digital camera, digitally reconstructed by a computer program, and the image is displayed on a monitor. The tale of holography starts in the upper left quadrant of the figure with the pioneering work of Gabor [15] of the first coherent optical hologram. Following Gabor's breakthrough, there are many interesting types of holograms in each of the four quadrants of Fig. 1, but owing to space limitations, we mention only the works that have been a direct inspiration for SLM-based imaging systems, which are the main topic of this review.

In the category of incoherent optical holograms, two seminal works published in 1966 introduced the principle of self-interference into the field of incoherent holography by recording an incoherent optical hologram with a triangular interferometer [16,17]. The self-interference principle indicates that the two interfering beams carry the object information, which is a conceptual leap from the Gabor hologram, in which the object information is carried out by only one of the interfering beams, whereas the other beam is used as a reference. In 1970, Bryngdahl and Lohmann added a theoretical analysis and more experiments to the topic of incoherent self-interference holography [18]. The principle of self-interference plays a significant role in SLM-based holographic systems, appearing more than 40 years after the first proposal of this principle. However, the number of publications in the field of optical incoherent holography in the twentieth century is much smaller than that in the other categories shown in Fig. 1, mainly because of a lack of appropriate technology, technical difficulties in implementing these systems, and the limited quality of the reconstructed images from such holograms. At the beginning of the twenty-first century, the general interest in optical holography, coherent or incoherent, declined in favor of digital holography [3].

The main activity in holography over the last thirty years, both in research and industry, has been in coherent digital holography, as shown in the lower left quadrant of Fig. 1. The digital hologram was introduced in 1967 by Goodman and Lawrence [19], and since then, this type of hologram has been the focus of many research groups [3]. In 1997, a phase-shifting procedure was proposed for on-axis digital hologram-recording setups [20] to address the twin-image problem [21]. A similar phase-shifting procedure is also used in some SLM-based holographic systems. However, most of the SLM-based holographic systems in this review belong to the category of incoherent digital holograms (IDHs) shown in the lower right quadrant of Fig. 1.

The first IDH is the innovative work of Poon, called optical scanning hologram (OSH) [22,23], but it is based on a different principle than self-interference. Instead, OSH uses a coherent heterodyne wave-interference pattern between two spherical waves that scans the observed three-dimensional (3D) object. The reflected light from the scanned object is collected by a point detector, enabling serial recording of a single IDH over time. This kind of 3D scanning contradicts the vision of the pioneers of holography to grasp the entire 3D scene in a single snapshot [15,24,25], and hence, this scanning encouraged the search for alternative IDHs. Thus, OSH was the main trigger to create FINCH in the sense that the FINCH inventors looked for a non-scanning hologram recording method, in contrast to using mechanical pixel-by-pixel scanning in 3D space as in OSH. On the other hand, the hologram recorded by OSH is of the Fresnel type [26], and in that sense, OSH inspired the first SLM-based FINCH system [14], since the inventors of FINCH also aspired to implement a Fresnel hologram, but in a different way.

FINCH, the first SLM-based system described in the next section, came to the world as a faster solution to the slowness problem of transverse and axial scanning of the input object in OSH and was inspired by some of the systems mentioned thus far, as indicated by the green arrows in Fig. 1. In other words, FINCH is an on-axis configuration such as Gabor's recording system [15], which makes use of the principle of self-interference proposed by Cochran and Peters [16,17] and analyzed by Bryngdahl-Lohmann [18]. Furthermore, FINCH uses the phase-shifting procedure proposed by Yamaguchi-Zhang [20], and it creates a Fresnel incoherent digital hologram as OSH does [22,23,26]. Another source of inspiration for FINCH, shown in the upper left quadrant of Fig. 1, is the groundbreaking optical correlator equipped with a Fourier hologram proposed in 1964 by Vander Lugt [27]. Since the FINCH system was supposed to record a Fresnel hologram, which is a two-dimensional (2D) correlation between the object function and a quadratic phase function, FINCH designers looked for an on-axis optical correlator to perform the task, such as the Vander Lugt correlator [27] (but adapted to incoherent illumination). Evidently, FINCH has its technological roots in numerous breakthroughs published over the past 60 years before its time, highlighting the importance of thoroughly studying the field and history of any invention before inventing it.

Fortunately, FINCH was not the end of SLM-based imaging systems but the beginning of a new conceptual era of unconventional optical 3D imaging that continues today with the help of many people worldwide. Section 2 is dedicated to FINCH and other closely related self-interference incoherent digital holograms. The system in which one of the SLM phase masks of FINCH is replaced from a diffractive spherical lens to an arbitrary scattering phase function is called a coded aperture holography (COACH) system, which is discussed in Section 3. Section 4 addresses another turning point in the development of SLM-based imaging systems, in which only one scattering phase mask has been displayed on the SLM, instead of multiplexing two as in COACH. Although this may seem like a minor technical change, displaying a single mask means that the object information is recorded without two-wave interference; hence, this topic is called interferenceless COACH (I-COACH). Surprisingly, the absence of two-wave interference has opened new applications, some of which are discussed in Section 4.1, under the topic of depth-of-field (DOF) engineering and optical sectioning. Section 5 is devoted to the discussion and conclusions of this review.

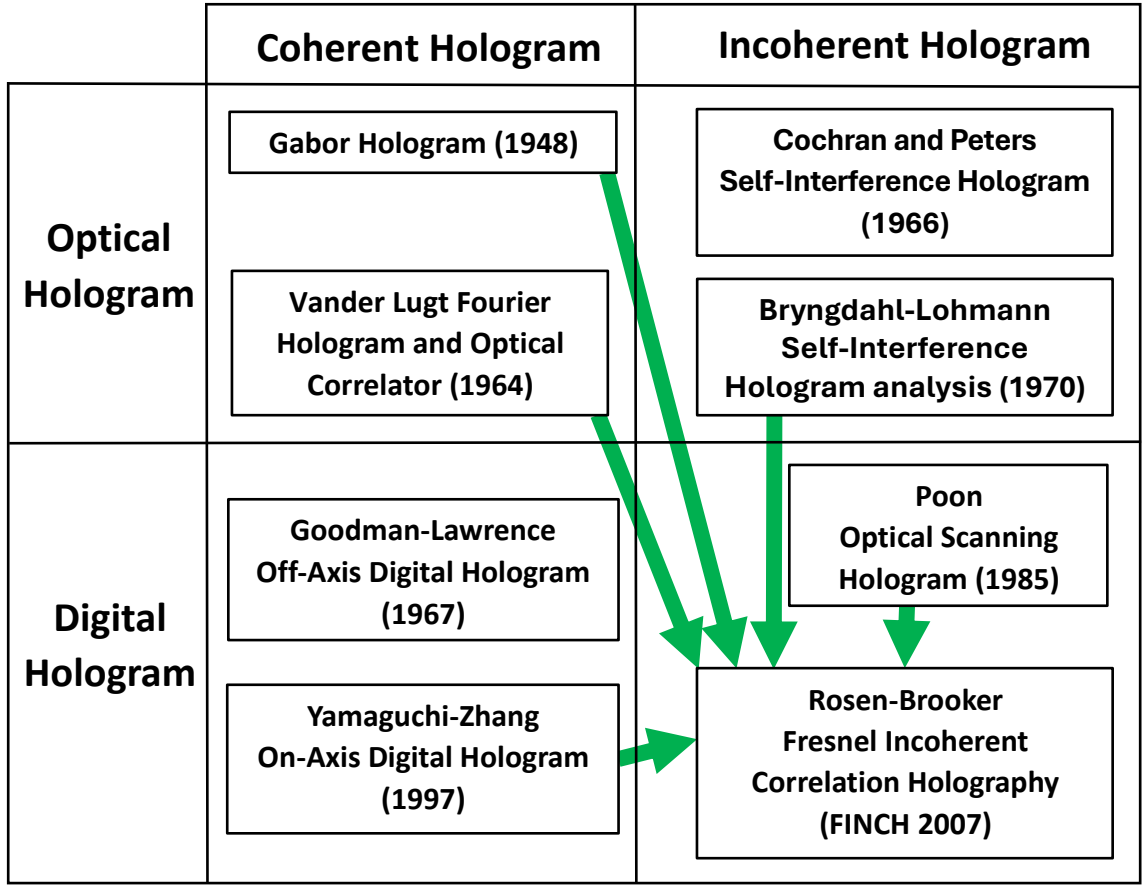

**Fig. 1** Scheme of the classification and main historical milestones of holographic imaging as described in the text. The green arrows indicate the influence of the various ideas on the Fresnel incoherent correlation holography (FINCH).

## 2. Fresnel and Fourier incoherent digital holography

The idea of FINCH with a phase-only SLM as its central component was initially proposed as a general approach to record self-interference incoherent digital holograms (SIDHs) [14] of 3D object scenes and was subsequently applied to fluorescence microscopy [28]. In retrospect, the process of inventing FINCH can be described by the block diagram of Fig. 2. The task was to implement a Fresnel hologram under spatially incoherent illumination. To accomplish this goal, two questions should have been answered. First, how to implement the mathematical operation for Fresnel propagation, namely, cross-correlation between an object function and a quadratic phase function, under incoherent illumination and in non-scanning mode, in contrast to the scanning mode of OSH [23]. The second question involved how to realize it in a single-channel incoherent correlator, in contrast to the well-known two-channel incoherent correlator [29]. The chosen replies to these two questions yielded the FINCH system shown in Fig. 3. In the first FINCH, a digital hologram was recorded by a single-channel setup with the help of the self-interference principle [16–18] and the flexibility of the phase-only SLM to multiplex more than one phase mask. Specifically, the light from each object point is split into two mutually coherent waves, where each wave is modulated by the SLM differently. Beyond the SLM, the interference of both waves with different wavefront curvatures is recorded by a digital camera. The interference pattern contains information about the 3D location and intensity of each point object, and this information can be decoded by a computer program. Because the system is linear, the process of imaging a single point can be generalized to imaging a complete object constructed from a continuous set of source intensity points.

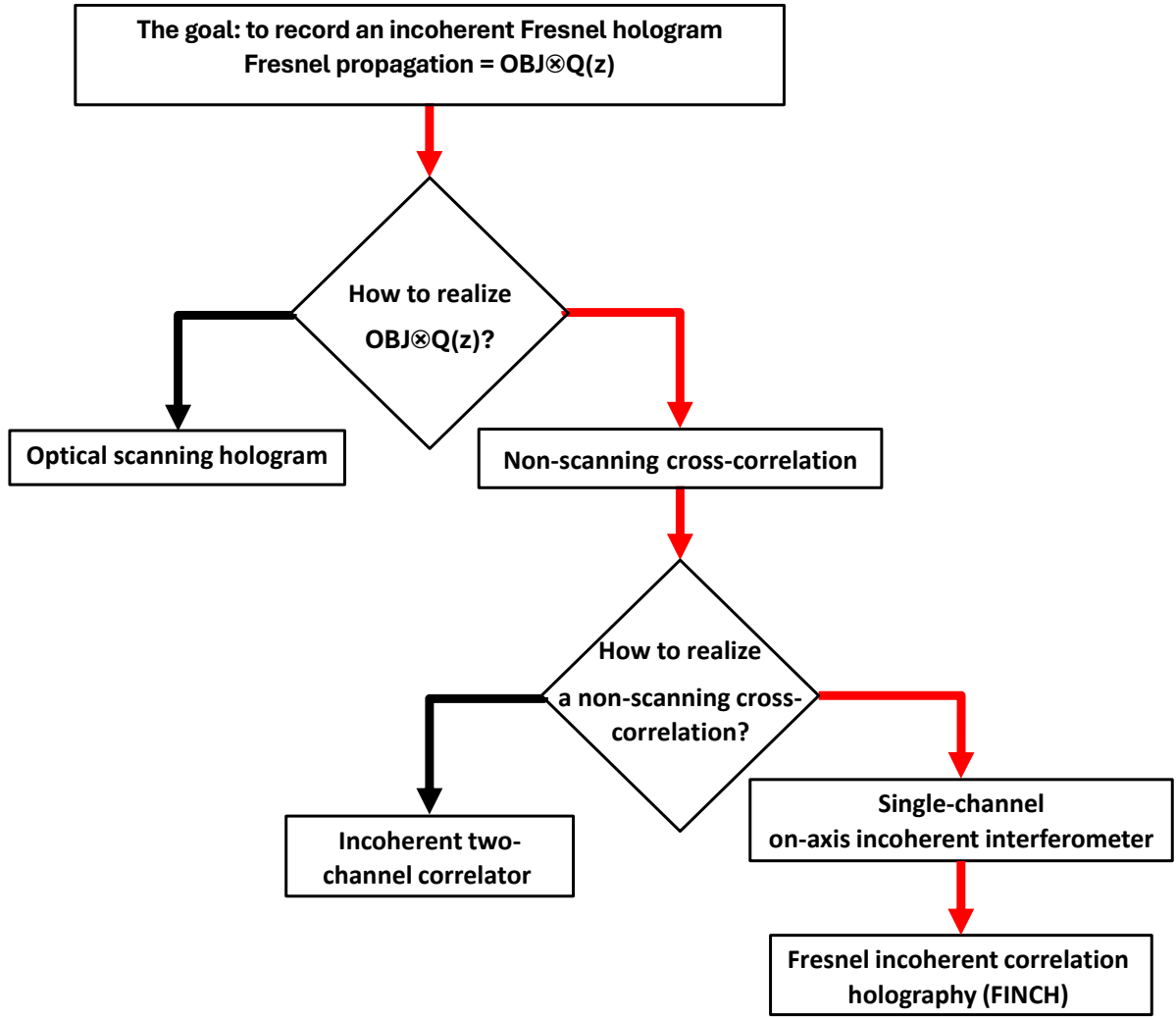

**Fig. 2** Scheme of the FINCH idea's birth in retrospect. The red arrows indicate the route that led to FINCH. The two solutions that have not been chosen are the optical scanning hologram (OSH) [23] and the incoherent two-channel correlator [29]. In the expression **OBJ⊗Q(z)**, **OBJ** is the object function, ⊗ denotes 2D correlation, and Q(z) is a quadratic phase function with the z parameter.

In FINCH, the phase-only SLM plays three different roles in the same device. First, the SLM operates as a beamsplitter such that the single wavefront emitted from the object point is split into two wavefronts by the two independent phase masks multiplexed on the SLM. Second, the light from each point source is focused toward two different focal points on the optical axis via the two diffractive lenses displayed on the SLM, where, in some cases, such as the one in Fig. 3, one focal point is at infinity. Third, the phase-shifting procedure [20] needed in on-axis digital holography to solve the twin image problem [21] can be enabled by multiplying only one of the diffractive lenses by at least three constant phases at three different times. Multiplexing the two diffractive lenses appeared in two versions. In the early years of FINCH, each element was displayed on half of the SLM pixels, with the pixel random partition between the two elements [14,28], as shown in Fig. 3. Since 2011 [30], the polarization sensitivity of the SLM has been used for multiplexing two masks because each orthogonal linear polarization is modulated differently; the wave polarized in the orientation of the SLM active axis is modulated by the single diffractive lens displayed on the SLM, and the wave polarized in the orthogonal orientation to the SLM axis is reflected unaffected by the SLM, as schematically illustrated in Fig. 4(a). This polarization multiplexing enables the recording of a single diffractive lens on all the SLM pixels; hence, the spatial noise of the first multiplexing method is significantly reduced [30].

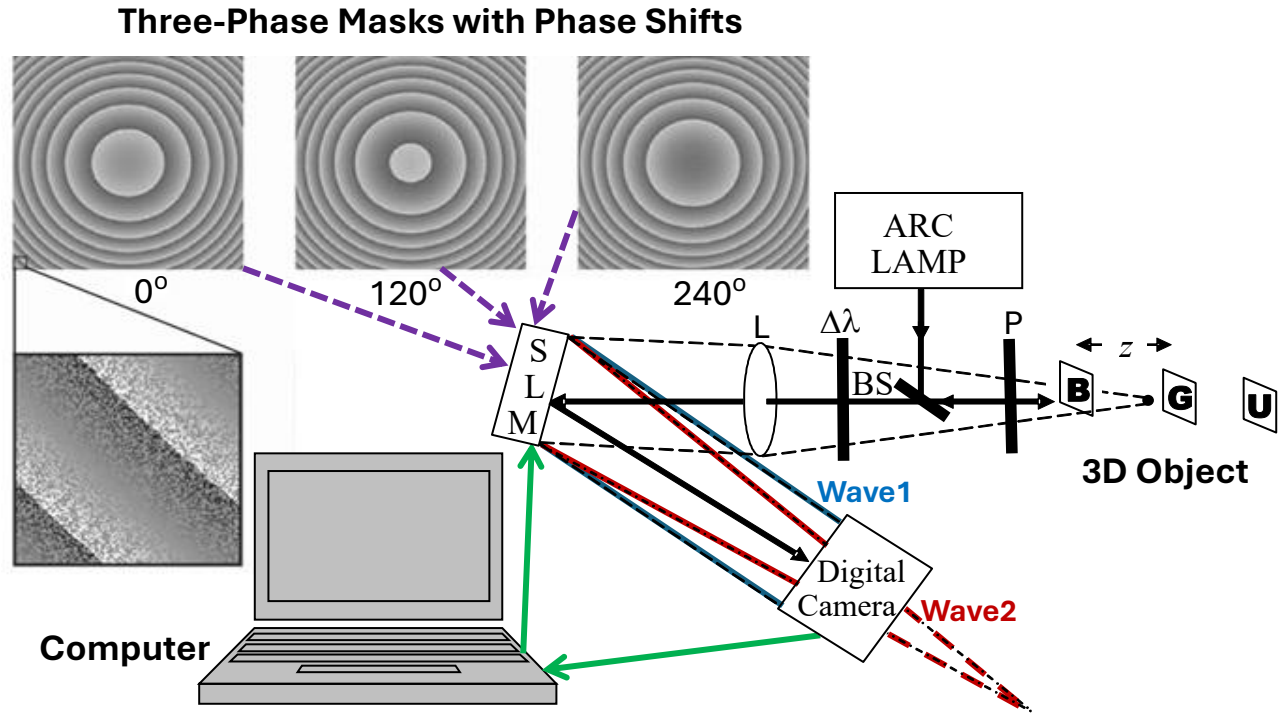

**Fig. 3** Scheme of the FINCH system with the early version of spatial multiplexing of two phase masks, a diffractive lens, and a zero phase mask. Spatial multiplexing is shown in the magnified part of the leftmost mask. SLM – spatial light modulators, BS – beamsplitter, L – refractive spherical lens, P – polarizer. The SLM is phase-only with polarization sensitivity in the orientation of the polarizer P.

As an optical imaging system, FINCH has unique features that distinguish it from other coherent or incoherent imaging systems. Under specific conditions, FINCH violates a general rule called the Lagrange invariant [31,32], which is satisfied by most optical imaging systems, in which the self-interference principle is not used. As a result of this violation, the lateral resolution of FINCH is superior to that of other coherent or incoherent systems with the same numerical aperture [32]. Further attempts to increase the transverse resolution have been made via different methods, such as structural illumination [33–35] and synthetic aperture [36–38].

FINCH has appeared in several versions proposed by different research groups [39–45], and some of these studies were summarized in a few review articles [46–49]. One notable development is a SIDH system with a single snapshot recording [50–67], which was recently proposed by several groups in several different versions to accelerate the recording process. Since FINCH is a type of SIDH, single-shot FINCH is included in the wider group of single-shot SIDH systems. The ways to achieve single-shot mode are diverse and include the following methods: A. Off-axis hologram recording [50,51]. B. Compressive sensing reconstruction [52–54]. C. Implementing parallel four-step phase-shifting holography by space-division multiplexing of polarization states using a polarization imaging camera [55–57]. D. Implementing parallel four-step phase-shifting holography by spatial multiplexing using multiplexed checkerboard phase gratings on two modified Michelson channels [58]. E. Nonlinear deconvolution using a library of point spread holograms (PSHs) [59]. F. Implementing parallel four-step [60] and two-step [61] phase-shifting holography along four and two optical axes, respectively, by spatial multiplexing of diffractive lenses on SLM. G. Implementing parallel three-step phase-shifting holography by space-division wavelength multiplexing [62]. H. Implementing parallel four-step phase-shifting holography by spatial multiplexing using a geometric phase-shifting grating [63]. I. Deep learning with neural networks [64,65]. J. Splitting the hologram by a mirrored phase-shifting module [66]. K. Hologram reconstruction with self-calibrated PSH [67]. All these systems significantly improve the temporal resolution and enable the capture of a video of a dynamic scene [66,68–70].

Another group of works that has become more widespread recently is FINCH systems supported by neural networks with deep learning, which were proposed to improve various qualities of images resulting from FINCH systems [71–73]. To date, most SIDH systems have contained no more than a single SLM, but several researchers have proposed two-SLM systems to gain various advantages. A Fourier incoherent holography system implemented in a single channel with two successive SLMs was proposed to reduce the optical path difference between the two interfering waves [74] in comparison to a single-SLM Fourier holography setup [75]. Since the Fourier incoherent holography systems also violate the Lagrange invariant, the inherent enhanced resolution mentioned above [32] is also valid for these systems. Setups of Fresnel SIDH with two sequential [76,77] and parallel [78,79] polarized-sensitive SLMs were proposed by Tahara. Such configurations save the use of polarizers and thus make the systems more power efficient.

FINCH and similar SIDH systems have been applied in several applications, most of which include fluorescence microscopy [28,30,32,62,71,80,81], incoherent microscopy [56,57,61,82–85], single-molecule localization [86,87], and lattice light-sheet microscopy [88–91]. Color 3D imaging by FINCH and similar systems has been demonstrated by several groups [56,92–99]. In these systems, the monochromatic SIDH was essentially extended to three channels, each of which is responsible for a different basic color. Optical sectioning of a 3D scene using FINCH was proposed in three versions; in two of them, sectioning is achieved via 3D pixel-by-pixel scanning, but the scanning type differs between versions. Electronic scanning by a second SLM was proposed in [100], whereas in [101], the second SLM was replaced by a polarizer with a pinhole in its center. Mechanical scanning in a similar way to a common confocal microscope was demonstrated in [102]. The third method of optical sectioning is based on a regular FINCH without scanning, and sectioning is achieved by a digital algorithm performed on the recorded holograms [103,104]. Extending the DOF is another notable application. A bimodal incoherent digital holography system was proposed in [105]. In one mode, regular three-dimensional imaging is achieved via a Fresnel digital hologram. In the second mode, the system can be turned into a recorder of digital Fourier holograms that yield quasi-infinite depth-of-field imaging. Recently, postrecording tuning of the depth of field has been successfully implemented in FINCH [106]. The interference conditions of the FINCH holograms were modified to create a library of FINCH holograms with different reconstruction planes and axial resolutions. These holograms are combined to extend the depth of focus of FINCH on-demand postrecording [106]. Edge enhancement in the reconstructed images was demonstrated by modifying the system's PSF to a vortex [107–109]. Phase recovery from Fresnel incoherent correlation holography was proposed using differential Zernike fitting [110]. A wide-view holographic stereogram was recently presented using multiple SIDH cameras [111]. A FINCH-based 3D ranging system was proposed in [112]. For cases of a phase aberrator between objects and a camera, Kim proposed a SIDH-based adaptive optics system where the image is reconstructed by a cross-correlation between the object hologram and guide-star hologram [113,114].

Notably, not all researchers have implemented SIDH with an SLM, although most of them acknowledged the influence of SLM-based FINCH on their work. Owing to the relatively high cost of the SLM and its pixelization, which induces diffraction noise, several research groups have preferred different ways of implementing self-interference digital hologram recorders without using SLMs. Some groups avoided using a single-channel SLM-aided interferometer and replaced it with an SIDH system implemented with a modified

Michelson interferometer [50,113–117], which splits the wave from each point object into two nonoverlapping wavefronts, each of which is modulated in a separate optical channel. Other research groups have preferred single-channel operation but have replaced SLMs with other devices, such as polarization-sensitive transmission liquid crystal gradient-index lenses [80]. Another example in this direction is a SIDH system, where the SLM was replaced with a bifocal isotropic metalens [53], a birefringent crystal lens [118], a bifocal geometric phase lens [54,98,119–122], or a successive combination of convex and half mirrors [123]. Another unusual modification in the conventional SIDH system is to replace the multipixel camera with a single-pixel camera and to use compressed imaging techniques to reconstruct the image [124]. The unavoidable conclusion from some of the above examples is that an SLM is a replaceable component that can be replaced with other components in incoherent hologram recorders. In the case of interferometers with two nonoverlapping channels, the SLM can be replaced by refractive elements such as spherical mirrors or glass lenses positioned differently in the two separated channels. In the case of a single channel, the SLM can be replaced by polarization-sensitive bifocal diffractive lenses. This is true as long as the mission is to record a self-interference IDH, and if the modulation of the light is performed via a bifocal mechanism. However, when one of the lenses from the bifocal elements is generalized to a chaotic diffractive phase element, the role of the SLM becomes more central, as discussed in the next section.

## 3. Coded aperture correlation holography (COACH)

Although the transverse resolution of FINCH is superior, its axial resolution is lower than that of other common imaging systems with the same numerical aperture. This disadvantage has encouraged the search for a different SIDH with a better axial resolution. Approximately one decade after the invention of FINCH, a new SIDH recorder, called COACH, with an axial resolution better than that of FINCH and equal to that of common direct imaging systems, was proposed [125]. In the COACH system, one of the modulating phase masks was generalized from a diffractive lens with a single focus of FINCH to a chaotic coded phase mask (CPM) with multiple focus points in COACH. The other phase mask multiplexed on the same SLM was a zero-phase mask, as in some versions of FINCH [14,28]. In COACH, an SLM was the prime component that could be replaced by a constant diffractive optical element [59] at the cost of losing the flexibility to change the CPM dynamically. In the COACH shown in Fig. 4(b), one of the two orthogonal polarizations is modulated by the CPM, whereas the other polarization propagates from an object to a camera unaffected by the CPM. The interference between the modulated and unmodulated beams originating from the same object point is recorded by the digital camera. As in the case of FINCH, the interference pattern contains information about the intensity and location of the source point of the two interfering waves. Unlike FINCH, however, the image cannot be reconstructed from the digital hologram via Fresnel backpropagation. Instead, a library of PSHs is prepared in advance, each of which is the system response to an object point located at a different $z$ location in the object space. To reconstruct the object's image in 3D space, three phase-shifted object holograms (OHs) are recorded once and processed via the phase shifting procedure as in the FINCH case. The obtained complex-valued hologram is processed by one of the available reconstruction algorithms [5] of 2D cross-correlation between the OH and each member of the PSH library. The sharpest image obtained from this process indicates the location of the original object in the 3D space. The differences between FINCH and COACH are schematically illustrated in Figs. 4(a) and 4(b).

After inventing the COACH, two studies were published. First, color sensing was integrated into COACH [126], and second, a hybrid FINCH-COACH system was proposed [127]. However, after a few months, researchers reported that for the task of 3D imaging, the optical interference between every two beams, both originating from every object point, is redundant. The intensity pattern recorded by a single wave modulated by the chaotic CPM provides complete information on the 3D location of the object. The self-interference principle that is essential in FINCH is no longer needed in COACH for 3D imaging. This difference between the systems is because the FINCH output waves are spherical waves, each with a constant magnitude, such that without self-interference, the 3D location of the source is lost. On the other hand, in COACH, the wave modulated by the scattering CPM has a chaotic magnitude that contains complete information about the 3D location of the point source. The intensity distribution recorded from a single wave by a camera can be decoded to reveal the intensity and the 3D location of the source. The system without two-beam interference is termed I-COACH [128], and it is the topic of the next section. Although COACH with wave interference has been revealed to be redundant for 3D imaging, there are several applications, such as quantitative phase imaging [129,130], multidimensional surface characterization [131], and synthetic aperture imaging [132], in which two-beam interference is still essential.

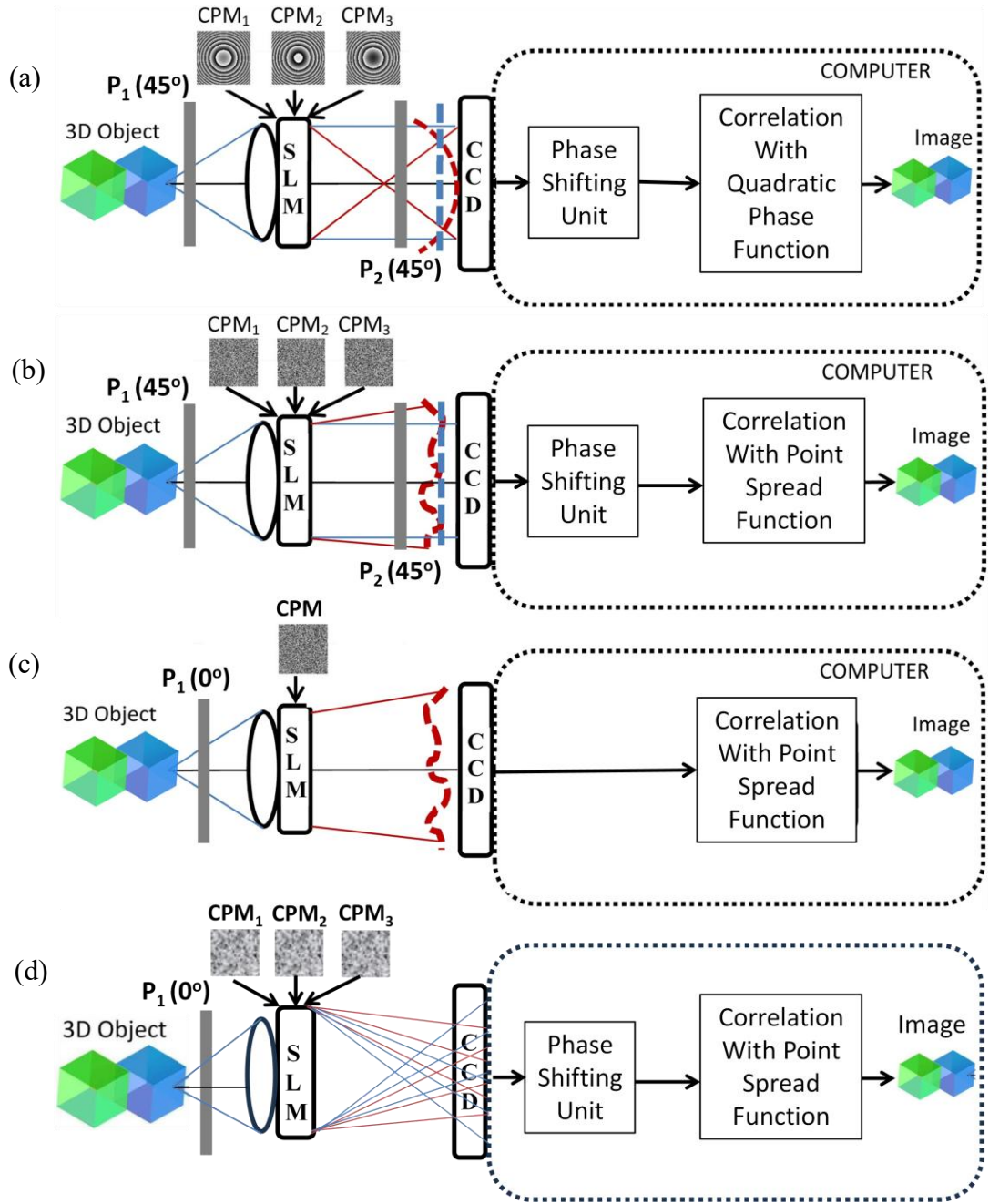

**Fig. 4** Schemes of the (a) FINCH, (b) COACH, (c) I-COACH, and (d) CAFIR systems. All the SLMs are phase-only with a polarization sensitivity in the orientation of 0°. CPM – Coded phase mask, SLM – Spatial light modulator, CCD – Charge–Coupled Device, $P_i$ – polarizer.

Another system based on two-wave interference is a combined COACH-FINCH scheme. Such a configuration improves lateral and axial resolution by leveraging the advantages of each system: FINCH's superior lateral resolution and COACH's improved axial resolution. The combination of FINCH and COACH to maximize the advantages of both is called the coded aperture with FINCH intensity response (CAFIR) [133]. CAFIR incorporates the superior axial resolution of COACH with the superior lateral resolution of FINCH. CAFIR can violate the Lagrange invariant, such as FINCH, in such a way that the transverse magnification of the gap between every two points is up to twice as large as the magnification of each point. This kind of violation is the reason for the enhanced lateral resolution of FINCH and CAFIR over regular COACH and direct imaging. In terms of axial resolution, the experimental results indicate that compared with FINCH, CAFIR preserves the axial resolution of direct imaging and regular COACH. The CAFIR system shows that sometimes, the combination of methods can yield a system with the best features of the combined methods instead of their mean features. CAFIR, created under the inspiration of FINCH, COACH, and I-COACH, as shown in Fig. 4(d), is evidence that different independent ideas can be combined into one system with superior features.

## 4. Interferenceless coded aperture correlation holography (I-COACH)

I-COACH, shown schematically in Fig. 4(c), has been investigated by many, from various aspects and for different applications [128,134–174]. The stages of development of I-COACH from FINCH through COACH are shown in Figs. 4(a)–4(c). If in COACH, one of the diffractive lenses of FINCH is replaced with a chaotic-scattering phase mask, in I-COACH, the scattering phase remains the only modulating mask of the SLM. Because of the single modulating mask on the SLM, the wave interference existing in FINCH of Fig. 4(a) and COACH of Fig. 4(b) is missing in I-COACH of Fig. 4(c). The images recorded by the digital camera in I-COACH are no longer holograms, at least not in terms of Gabor's meaning, since no wave interference is involved in their creation. Hence, in the case of a point input, the intensity system response is called the point spread function (PSF), and if a multipoint object is introduced into the I-COACH system, the intensity response is termed the system-to-object response (SOR). The I-COACH image is reconstructed via algorithms in which the SOR is cross-correlated with the PSF library. The number of camera shots has also undergone several developments, and the most recent examples present single-shot systems. Because of the lack of wave interference, the phase-shifting procedure is not needed. Moreover, because of the use of advanced linear and nonlinear reconstruction algorithms, the process of SOR acquisition becomes a single-shot process [135,152,153,162,163,175]. Single-shot recording means that only one CPM is displayed on the SLM, as shown in Fig. 4(c). In principle, a static phase-only diffractive optical element can replace the dynamic SLM, but at least in the stage of research and CPM design, the SLM provides the flexibility to test many different CPMs. Moreover, using SLM instead of a static phase mask makes the system flexible enough to move from one application to another just by changing the CPM displayed on the SLM electronically, whereas replacing a diffractive optical element can be done only mechanically.

Since the I-COACH invention in 2017 [128], several researchers have proposed improvements, mainly in terms of image reconstruction methods [141,147,153,159,163,164,176] and PSF engineering. Following the first I-COACH, the PSFs in the early years were quasirandom speckle distributions limited by some rectangles [128,141,147,159,163]. Some groups have suggested a ring-shaped PSF [148,156]. Other groups investigated I-COACH systems with PSFs of longitudinally structured beams [152,164,177]. In some other studies [139,175], the PSF in the form of a dot pattern was implemented because it has a higher signal-to-noise ratio (SNR) than other PSFs do. The disadvantage of dot-shaped PSFs is that the dot pattern is obtained only in a single plane along $z$. For 3D imaging, several CPMs should be multiplexed, where each CPM is responsible for a different dot pattern at a different depth [175].

I-COACH has been applied to several applications in many ways, all of which rely on a phase-only SLM. Among the several applications that are worth mentioning are 3D imaging with the ability to focus each time at a different lateral plane [175], 3D color imaging [177], extending the field of view [136,178], 3D imaging through scatterers [143], depth-of-field engineering [145], optical sectioning [179], polarization imaging [171], capturing a video of a dynamic scene [177], and improving the lateral resolution [142,160,166,173], as discussed next.

The connection between holography and resolution enhancement has been well known for many years [180]. As mentioned in Section 2, FINCH has been found to have a unique feature that makes this system superior in terms of lateral resolution over other imaging systems with the same numerical aperture. COACH and I-COACH lack this superiority, and they both satisfy the Lagrange invariant rule, which gives them a lateral resolution similar to that of other conventional imaging systems. Integrating SLMs into the I-COACH configuration can make such systems dynamic and add a time dimension to the three spatial dimensions of the various holographic systems. For example, our group recently proposed a method for improving the image resolution of diffraction-limited imaging systems [181]. The resolution is enhanced by a deconvolution process, and unavoidable noise, which appears because of the deconvolution, is averaged down by multiple image acquisitions, where each image is recorded with a different phase aperture displayed on the same SLM. However, methods based on changing CPMs over time, such as those proposed in [181], do not extend the effective numerical aperture; therefore, the resolution improvement is limited. Other I-COACH-based methods in which the effective numerical aperture is extended have been recently proposed, including a scattering phase mask between the object and the system's aperture [142,160], synthetic aperture imaging [132], and imaging with structured light illumination [173].

## 4.1 Depth of field engineering and optical sectioning by I-COACH

Extending the DOF of an optical imaging system without affecting other imaging properties has long been an important topic of research. A recently developed area beyond the simple extension of the DOF is DOF engineering, which is achieved with the help of phase-only SLMs [145,182]. DOF engineering means that the characteristic DOF of an imaging system can be extended to one or several separate axial intervals with different lengths. Each interval has controlled start and end points, and the overall DOF engineering occurs without affecting the lateral image resolution. Because of the DOF engineering, objects in certain separate, different input subvolumes are imaged with the same sharpness as if these objects were all in focus. Additionally, the images from different subvolumes can be laterally shifted, with each subvolume shifting differently relative to its original position in the object space. By performing lateral shifts, mutual hiding of images can be avoided. The proposed method was demonstrated on an I-COACH system with a multidot PSF [145]. This means that the light from the object space is modulated by a CPM and recorded into the computer, in which the desired image is reconstructed from the recorded pattern. The DOF engineering was achieved by designing a phase mask composed of three diffractive elements, which was displayed on the SLM. One mask was a quadratic phase function, which is a diffractive lens, dictating the starting point of the in-focus axial interval, and the second mask was a quartic phase function, which dictates the end point of this interval. The CPM for creating a PSF of several randomly distributed dots was the third diffractive element, which enables digital image reconstruction via deconvolution between the SOR and PSF. Multiplexing several sets of three diffractive elements, each with a different set of phase coefficients, can yield various DOF intervals along the $z$-axis. All the diffractive elements are displayed on a spatial light modulator such that real-time DOF engineering is enabled according to the user's needs during the observation.

A recent development in the field of DOF engineering is a multipurpose system in which, for the same single-shot SOR, one can reconstruct a single transverse plane or multiple planes of images from any combination of subspaces along the $z$-axis in the computer [175]. This technique has enabled the reconstruction of only a specific transverse plane or multiple planes at a time from the same single-shot recorded SOR, and, according to the user's wishes, by postprocessing this SOR in the computer. An SOR is recorded while keeping $N$ object subspaces in focus with a phase mask introduced in the system's aperture. The phase mask is a multiplexing of $N$ sets of unique scattering phases multiplied by unique quadratic phases (diffractive lenses) such that each set has a unique focal length. Each set is assigned to a specific subspace of the object space. Each scattering phase is intended to yield a unique pattern of randomly distributed dots on the camera. Any of the $n$th ($1 \leq n \leq N$) subspaces can be reconstructed by deconvolving the SOR with the corresponding pattern of random dots. Simultaneous reconstruction of $1 \leq n \leq N$ planes in the volume can be obtained by deconvolving SOR with the combined pattern of random dots. The multiple features of this single-shot technique make it cost-effective and time-efficient, as different aspects of the observed scene can be reconstructed from a single SOR without the need to record another. The 3D PSF of the system for $N=3$ and $K=3$ dots is shown in Fig. 5. Each object located in the $n$th subspace can be reconstructed by the $n$th PSF of $K$ dots, which are arranged randomly and differently for each $n$th subspace.

Imaging thick and complex objects remains a significant challenge because out-of-focus information produces blurred background noise. Confocal imaging [102] offers a fully noninvasive alternative, enabling the acquisition of high-contrast 2D image stacks from 3D specimens with minimal background noise. However, current state-of-the-art confocal systems depend on pixel-by-pixel scanning, which substantially increases both the imaging time and system cost. Recently, a new non-scanning optical method for sectioning and background noise suppression was introduced based on I-COACH [179]. This technique is called sectioning longitudinal images via a complex-correlation equation (SLICE). SLICE is based on the same system with the same PSF shown in Fig. 5. The new principle of SLICE is the use of three camera recordings that are digitally projected onto the complex space; each camera recording is the system's response to the appearance of a multiplane object at the system's input. The three camera recordings differ from each other since the object light is modulated by three different CPMs, each of which creates a random and different set of dots. Image reconstruction is achieved through deconvolution between two complex-valued functions, the SOR and the computer-generated PSF. The phase distribution of the deconvolution serves as a filter enabling reconstruction of images in the desired $z$-slice only, excluding out-of-focus images and minimizing background noise.

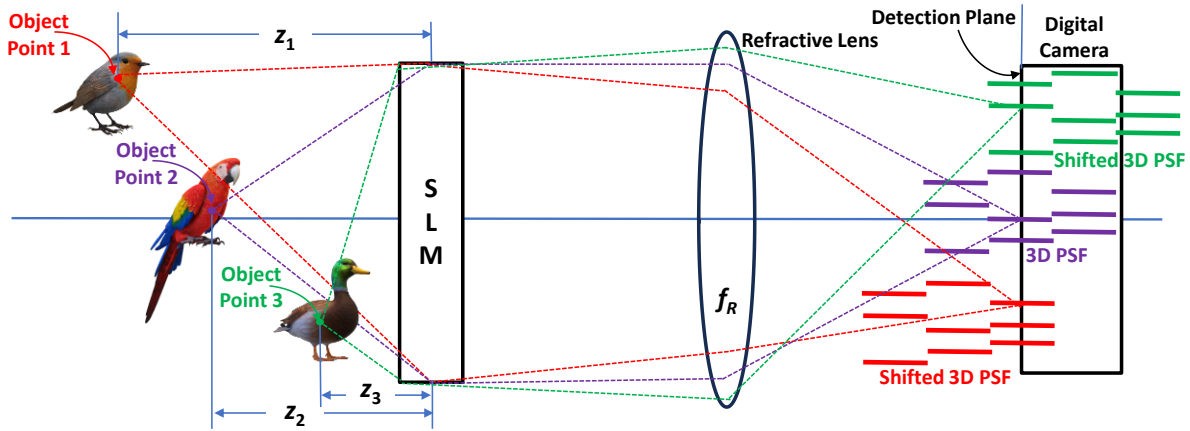

**Fig. 5** Schematic of the SLICE system. The 3D PSF is obtained in response to object point 2, whereas the shifted 3D PSFs are obtained in response to object points 1 and 3. SLM – spatial light modulator, RL – refractive lens, $f_R$ is the focal length of the RL, PSF – point spread function. The horizontal lines of the 3D PSF indicate the depth of focus of the system.

## 5. Summary and conclusion

The tale of the SLM-based 3D imaging systems in the last two decades is an example of how the development of one device, the SLM in this case, can trigger the development of other nonrelated technologies, such as 3D imaging systems. The evolution of these imaging systems is a story of the successful combination of two unrelated technologies. On the one hand, phase-only SLMs appeared at the end of the 1980s and were used mainly as electro-optical displays for projecting images in the air or on a screen. On the other hand, holography researchers looked for an efficient way to record IDHs without pixel-by-pixel scanning. As described in this article, the integration of phase-only SLMs into the technology of 3D imaging systems has solved various problems of optical imaging and has been implemented in various systems for diverse imaging applications.

Moreover, the addition of SLMs into imaging systems can yield new technologies with superior capabilities over other technologies, such as axial sectioning without scanning and DOF engineering discussed in subsection 4.1.

Nonetheless, this SLM-based imaging technology should still be improved in terms of hardware and software. Possible applications that can benefit from this technology should be sought. In terms of hardware, better SLMs with more and smaller pixels undoubtedly improve imaging performance. Another aspect of hardware improvement is the need to produce SLMs in transmission mode rather than just reflective mode. This change can make optical systems more compact, aligned with a single optical channel, and more energy efficient by avoiding the use of unnecessary beamsplitters. From the software aspect, more sophisticated algorithms for image reconstruction are expected to increase the quality of the output images. With respect to the possible applications that can be implemented by this technology, topics such as imaging through a scattering medium, 3D imaging in spectral regions beyond the visible wavelengths, and quantitative phase imaging will likely benefit from this technology in the near future, and initial attempts in these directions have already been made. Most examples of practical applications demonstrating the real-world utility of these techniques exist in the microscopy field. The proposed techniques surveyed in this article provide the capabilities of 3D imaging and axial sectioning to new microscopes. Based on many research articles on these topics, the tale of SLM-aided imaging systems is far from ending, and several new SLM-based imaging systems will probably appear in the coming years.

The SLM-based imaging systems discussed in this review are the result of the technological evolution of optical holography, spanning almost 80 years. It started in 1948 with the invention of the revolutionary Gabor hologram, in which one of the interferometer beams carried a wavefront that diffracted from the observed object, and the other beam was used as a reference. However, Gabor's concept of object and reference beams cannot be applied to holograms of objects that emit spatially incoherent light because wave interference between the object and reference beams is impossible under incoherent light. The solution to the problem of incoherence between the two waves involves the principle of self-interference, in which each of the interfering beams is coherent with the other because both originate from the same source point, and each beam carries a different image of the same object. FINCH with the SLM enables the combination of self-interference with the phase-shifting procedure to record IDHs in a single-channel setup. The next development stage, COACH, was introduced as an additional conceptual progress in which the image of the object is replicated and randomly distributed in the camera space by one of the interfering beams, whereas the other beam carries a single image as before. While the evolution of 3D holographic recorders became increasingly complicated, the I-COACH appeared surprising, with the demonstrated claim that two-beam interference is not needed at all, at least not for 3D imaging; hence, only one beam from the object point remains in I-COACH and is scattered by the CPM. For other applications, such as quantitative phase imaging, synthetic aperture, and other superresolution techniques, two-beam interference still plays an important role; hence, some configurations of COACH are still helpful. In both the COACH and I-COACH configurations, the use of phase-only SLMs is extensive, enabling the development of new systems for new applications.

The development from FINCH to COACH and from COACH to I-COACH is not incidental. Any innovation starts with the motivation of its inventors. Problems with the existing system, such as being slow or inefficient, usually trigger the inventors to solve these problems with new development. The motivations of FINCH, COACH, and I-COACH are different. FINCH was proposed as a solution to the problem of transverse and axial scanning of the input object in OSH. However, the main problem of FINCH is its low axial resolution, which is attributed to the CPM of FINCH, the quadratic phase function. This FINCH's problem defines the main motivation behind COACH—to improve the axial resolution of FINCH without reducing the lateral resolution below that of lens-based systems. However, the weakness of FINCH and COACH is the need for two-beam interference to record an IDH of the 3D scene. At this point, the motivation to replace COACH with an interferenceless system led to the proposal of I-COACH as a system that can provide the same 3D imaging capabilities without two-beam interference.

Once the motivation is clear, start the stage of analyzing the existing system with its inherent weaknesses to fully understand its operation. When this operation is understood, the inventors search for a system that can perform the same operation and does not suffer from the same troubling problem. The search can be based on the personal knowledge and experience of the inventors, or it can demand a thorough investigation into the relevant literature. Usually, the basic ideas of innovation exist in many, sometimes nonrelated systems. Fig. 1 is a graphical demonstration of the FINCH invention as an example of several ideas existing in a few systems that have been combined into a new system. In short, an invention is usually a new way of combining old ideas. Therefore, my main lesson from thirty years of working on 3D imaging systems, some of which are described in this article, is that usually inventing something new is about collecting old ideas that exist in several different systems and combining them, but in a new way. The creative talent of people should lead them to find an original way to combine all these known ideas into something new and useful. Ironically, even this lesson is not completely new but is similar to a more than 100-year-old quote of the famous American writer Mark Twain, who used a nice metaphor from the world of optics, saying that "There is no such thing as a new idea. It is impossible. We simply take a lot of old ideas and put them into a sort of mental kaleidoscope. We give them a turn, and they make new and curious combinations. We keep on turning and making new combinations indefinitely; but they are the same old pieces of colored glass that have been in use through all the ages" [183]. This tangible and insightful quote is additional support for the comment in the Introduction that it is important to know as many existing ideas as possible to be able to combine some of them in new ways, or, in Winston Churchill's words, "The farther backward you can look, the farther forward you are likely to see" [184]. This review attempts to make a modest contribution to expanding knowledge, at least in holography and optical imaging.

**Funding** Supports are acknowledged from the Israel Science Foundation (ISF) Grant No. 3306/25.